\documentclass[12pt]{article}
\usepackage{amsmath,amssymb}
\usepackage{bm}
\usepackage{color}
\usepackage{cite}
\usepackage{mathtools}

\usepackage{multirow}

\def\slash#1{\not\!\!#1}

\begin{document}

\title{
\begin{flushright}
\ \\*[-80pt]
\begin{minipage}{0.25\linewidth}
\normalsize
EPHOU-23-003\\
KYUSHU-HET-255 \\*[50pt]
\end{minipage}
\end{flushright}
{\Large \bf
Remark on modular weights in low-energy \\ effective field theory from type II string theory
\\*[20pt]}}

\author{
Shota Kikuchi$^{1}$,
Tatsuo Kobayashi$^{1}$,
Kaito Nasu$^{1}$, \\
 Hajime Otsuka$^{2}$, 
Shohei Takada$^{1}$, 
and
~Hikaru Uchida$^{1}$
\\*[20pt]
\centerline{
\begin{minipage}{\linewidth}
\begin{center}
$^{1}${\it \normalsize
Department of Physics, Hokkaido University, Sapporo 060-0810, Japan} \\*[5pt]
$^{2}${\it \normalsize
Department of Physics, Kyushu University, 744 Motooka, Nishi-ku, Fukuoka 819-0395, Japan} \\*[5pt]
\end{center}
\end{minipage}}
\\*[50pt]}

\date{
\centerline{\small \bf Abstract}
\begin{minipage}{0.9\linewidth}
\medskip
\medskip
\small
We revisit the modular weights in type IIB magnetized D-brane models.
The simple analysis of wave function shows that the four-dimensional matter fields have 
the modular weight --1/2, but it may shift as one in 
type IIA intersecting D-brane models.
For example, the localized gauge flux as well as the localized curvature can 
shift the modular weight in the magnetized D-brane models.
Such corrections do not affect physical couplings such as physical Yukawa couplings.
However, it leads to differences in supersymmetry breaking sfermion masses, which depend on the modular weights, 
although the $A$-term coefficients and the sum of sfermion masses squared seem to be 
the same between two models.
\end{minipage}
}

\begin{titlepage}
\maketitle
\thispagestyle{empty}
\end{titlepage}

\newpage


\section{Introduction}
\label{Intro}

Modular symmetry is a geometrical symmetry of torus and its orbifolds.
Moduli, which parametrize geometrical aspects of these compact spaces such as 
size and shape, transform non-trivially under the modular symmetry.
Calabi-Yau manifolds have many moduli and 
larger modular symmetries.
These modular symmetries appear in four-dimensional (4D) low-energy effective 
field theory derived from the string theory.
(See for 4D effective field theories from the string theory and their phenomenological aspects Refs.~\cite{Ibanez:2012zz,Blumenhagen:2006ci}.)
Yukawa couplings and other couplings in such 4D effective field theories depend on 
moduli, and they transform non-trivially under the modular symmetry.
Chiral matter fields also transform non-trivially under the modular symmetry \cite{Ferrara:1989bc,Ferrara:1989qb,Lerche:1989cs,Lauer:1989ax,Lauer:1990tm,Kobayashi:2016ovu,Kobayashi:2018rad,Kobayashi:2018bff,Ohki:2020bpo,Kikuchi:2020frp,Kikuchi:2020nxn,
Kikuchi:2021ogn,Almumin:2021fbk,Baur:2019iai,Nilles:2020kgo,Baur:2020jwc,Nilles:2020gvu}.
Thus, the modular symmetry is in a sense the flavor symmetry\footnote{See for larger symplectic modular symmetries on Calabi-Yau compactifications Refs.~\cite{Strominger:1990pd,Candelas:1990pi,Ishiguro:2020nuf,Ishiguro:2021ccl}.}.

The modular group includes finite subgroups $S_3$, $A_4$, $S_4$, and $A_5$ \cite{deAdelhartToorop:2011re}.
These discrete symmetries have been used 
as flavor symmetries in a bottom-up approach of flavor model building \cite{
	Altarelli:2010gt,Ishimori:2010au,Kobayashi:2022moq,Hernandez:2012ra,
	King:2013eh,King:2014nza}.
Inspired by these aspects, modular flavor symmetric models have been studied extensively in both top-down and 
bottom-up approaches\footnote{
See for earlier works Refs.~\cite{Feruglio:2017spp,Kobayashi:2018vbk,Penedo:2018nmg,Criado:2018thu,Kobayashi:2018scp,Novichkov:2018ovf,Novichkov:2018nkm,deAnda:2018ecu,Okada:2018yrn,Kobayashi:2018wkl,Novichkov:2018yse}
and see for more references \cite{Kobayashi:2022moq}.}.

In modular symmetric field theories, modular weights of chiral matter fields are important quantities.
Certain modular weights of matter fields appear in 4D effective field theories derived from the string theory.
In heterotic string theory, 4D low-energy effective field theory can be studied by 
dimensional reduction \cite{Witten:1985xb}.
Untwisted matter fields on orbifolds and Calabi-Yau compactifications have 
the modular weight $-1$.
Furthermore, the modular weights of twisted sectors on $Z_N$ orbifolds were obtained through string amplitude computations 
by use of conformal field theory \cite{Dixon:1989fj,Kaplunovsky:1995jw}.
(See also Refs.~\cite{Ibanez:1992hc,Ibanez:2012zz}.)
Those modular weights are fractional numbers depending on twist angles of twisted strings $\theta=k/N$.

Similarly, in type IIA intersecting D-brane models, 
modular weights of 4D matter fields were studied in 
\cite{Blumenhagen:2006ci,Kors:2003wf,Lust:2004cx,Akerblom:2007uc,Blumenhagen:2007ip}, and they are fractional numbers related with intersecting angles $\theta_{ab}$ between two D-branes, 
D$_a$ and D$_b$.
For example, in Refs.~\cite{Akerblom:2007uc,Blumenhagen:2007ip}, 
one-loop threshold corrections to gauge couplings were studied.
A reasonable Ansatz on the modular weights was proposed in type IIA intersecting D-brane models.
That is, the modular weight can be written by $k=-1/2\pm \theta_{ab}/2$, which is consistent with 
one-loop gauge couplings.

In type IIB magnetized D-brane models, 4D low-energy effective field theory 
was studied through the dimensional reduction, where non-trivial wave functions of matter fields 
with background magnetic fluxes were used \cite{Cremades:2004wa}.
Those analyses on wave functions show that 
for $T^2$ and its orbifolds, chiral matter fields have the modular weight $-1/2$ \cite{
Ohki:2020bpo,Kikuchi:2020frp,Kikuchi:2020nxn,
Kikuchi:2021ogn}.
However, these modular weights in magnetized D-brane models may have corrections 
like heterotic orbifold models and intersecting D-brane models.
In this paper, we study such corrections and examine their implications.

This paper is organized as follows.
In section \ref{sec:review}, we briefly review the modular weights of chiral matter zero-modes in 
type IIB string theory with magnetized D-branes. In section \ref{sec:LEEFT}, the implications of corrections to the modular weights are discussed in the context of low-energy effective field theory. We examine the modular weight corrections in type IIA intersecting D-brane models and type IIB magnetized D-brane models in sections \ref{sec:typeIIA} and \ref{sec:singular}, respectively. In section \ref{sec:SUSY}, we study the modular weight corrections to soft supersymmetry-breaking terms. The modular weights of oscillating modes and massive modes are discussed in section \ref{sec:massive}. 
Section \ref{sec:Conclusion} is devoted to the conclusion. 
In Appendices \ref{app:wave} and \ref{app:modular_weight}, we summarize matter wave functions and their modular weights, respectively.

\section{Modular weights of chiral matter fields}
\label{sec:review}

Here, we briefly review the modular weights of chiral matter fields in type IIB magnetized D-brane models.
For simplicity, we consider the six-dimensional compact space, which is factorizable as 
$T^6=T^2\times T^2 \times T^2$ and its orbifolds.
We concentrate one $T^2$ among $T^2\times T^2 \times T^2$.
We denote the complex coordinate on $T^2$ by $z$ and the complex structure modulus by $\tau$.
For simplicity, we consider the $U(1)$ gauge theory with the following magnetic flux background,
\begin{align}
F= \frac{\pi i M}{{\rm Im} \tau} dz \wedge d\bar z.
\label{eq:magne}
\end{align}
Here, $M$ must be an integer because of Dirac's quantization condition.
Under this magnetic flux background, the Dirac equation for the fermion with $U(1)$ charge $q=1$ 
has zero-modes, whose number is $|M|$.
As shown in Appendix \ref{app:wave}, 
for $M>0$, their wave functions can be written by \cite{Cremades:2004wa}
\begin{align}
\psi^{j,M}(z,\tau) = \frac{M^{1/4}}{{\cal A}^{1/2}}  e^{\pi iMz\frac{{\rm Im}z}{{\rm Im}\tau}}
  \vartheta \begin{bmatrix}
    \frac{j}{M} \\
    0 \\
  \end{bmatrix}
  (Mz,M\tau), \label{eq:psi} 
\end{align}
where the index $j$ ($j=0,1,\cdots, M-1$) labels degenerate zero-modes, and  
${\cal A}$ is the area of $T^2$, and $\vartheta$ denotes the Jacobi theta function defined by
\begin{align}
  \vartheta
  \begin{bmatrix}
    a \\ b \\
  \end{bmatrix}
  (\nu,\tau)
  = \sum_{\ell\in\mathbb{Z}} e^{\pi i(a+\ell)^2\tau} e^{2\pi i(a+\ell)(\nu+b)}.
\label{eq:theta-f}
\end{align}
Similarly, we can obtain the zero-mode wave functions when $M<0$.

Because of modular transformation behavior\cite{Cremades:2004wa,Kobayashi:2018rad,Kobayashi:2018bff,
Ohki:2020bpo,Kikuchi:2020frp,Kikuchi:2020nxn,
Kikuchi:2021ogn}, it would be natural to normalize the 
wave functions as 
\begin{align}
  &\int d^2z ~\psi^{i,M}(z,\tau) \left(\psi^{j,M} (z,\tau)\right)^* = (2{\rm Im}\tau)^{-1/2}\delta_{i,j}. \label{eq:NormalizationPSI}
\end{align}
Indeed, under the $S$-transformation of the modular symmetry, 
\begin{align}
S:~\tau \to -1/\tau, \qquad z \to -z/\tau,
\end{align}
the wave functions transform 
\begin{align}
\psi^{j,M}(z,\tau) \to (-\tau)^{1/2} \sum_k \rho(S)_{jk} \psi^{k,M}(z,\tau),
\end{align}
where $\rho(S)_{{jk}}=M^{-1/2}e^{i \pi /4}e^{2\pi i\,jk/M}$.
Thus, the wave functions have the modular weight $1/2$.

Orbifold compactification is also one of the interesting compactifications.
The $T^2/Z_N$ orbifold is constructed by imposing $Z_N$ identification on $T^2$, i.e., 
$\theta z \sim z$ on the $T^2$ coordinate $z$, where $\theta^N=1$.
Wave functions on the $T^2/Z_N$ orbifold have twisted boundary condition, 
\begin{align}
\psi^{j,M}_m(\theta z,\tau) = e^{2\pi i m/N}\psi_m^{j,M}(z,\tau).
\label{eq:twist-BC}
\end{align}
Wave functions on the orbifold are constructed by those on $T^2$ as \cite{Abe:2008fi,Abe:2013bca}
\begin{align}
\psi^{j,M}_m(z,\tau) = {\cal N} \sum_k \psi^{j,M}(\theta^kz,\tau)e^{-2\pi ikm/N},
\end{align}
where ${\cal N}$ is a normalization constant.
The degeneracy number of zero-modes depends on the eigenvalue of the $Z_N$ twist as well as $M$ \cite{Abe:2008fi,Abe:2013bca}.

We assume that 4D $N=1$ supersymmetry (SUSY) is preserved.
Then, there appear massless scalar fields, which are superpartners of the above fermion, 
and they have the same wave functions with the same degeneracy.
For simplicity, we start with six-dimensional (6D) field theory on $T^2$
instead of ten-dimensional (10D) theory on $T^2\times T^2 \times T^2$
and consider the bosonic and fermionic fields $\Phi(x,z)$ and $\Psi(x,z)$, where $x$ denotes 4D spacetime coordinates.
We write them by Kaluza-Klein decomposition:
\begin{align}
&\Phi(x,z)=\sum_j\phi_j(x)\psi^{j,M}(z,\tau) + \cdots, \notag \\
&\Psi(x,z)=\sum_j\chi_j(x)\psi^{j,M}(z,\tau) +\cdots,
\end{align}
where the first terms correspond to massless modes and the ellipsis denotes massive modes.
Here, we have omitted the spinor index.
We further assume the canonical kinetic term 
\begin{align}
\partial_{\cal M} \Phi^* \partial^{\cal M} \Phi,
\end{align}
in the 6D theory with ${\cal M}=0,1,\cdots,5$.
Then, we carry out the dimensional reduction by integrating the $z$ coordinate and using the above normalization 
Eq.~(\ref{eq:NormalizationPSI}) so as to find 
the 4D kinetic terms, corresponding to the K\"ahler potential,
\begin{align}
K=K_{i \bar i}|\phi_i(x)|^2=\frac{1}{(2 {\rm Im} \tau)^{1/2}}|\phi_i(x)|^2.
\end{align}
Thus, the matter fields have the modular weight $k= -1/2$.
Similarly, we can compute the fermionic modes and they have the same modular weight.
Also, the computation on the $T^2/Z_N$ orbifold is the same.
Furthermore, the extension to 10D theory on $T^2\times T^2 \times T^2$ and orbifolds 
is straightforward.
Chiral matter fields have the modular weights  $-1/2$ for the modulus $\tau_\alpha$ on the $\alpha$-th $T^2$.
This result is universal and independent of the magnetic flux.
Note that if the wave function in the compact space has the modular weight $k$, 4D matter field has 
the modular weight $-k$.

\section{Low-energy effective field theory}
\label{sec:LEEFT}

The modular weights of matter fields may have corrections, that is,  shifts from the weight $k=-1/2$ in magnetized D-brane models.
Here, we study the implications of the modular weight shifts in low-energy effective field theories. 
The K\"ahler metric gives us the normalization of kinetic term, and 
such a normalization is not physically observable.
The important point is that physical 3-point couplings are normalized by  the factor $(K_{i\bar i}K_{j \bar j}K_{k\bar k})^{-1/2}$, 
where $K_{i \bar i}$, $K_{j \bar j}$, and $K_{k \bar k}$ are K\"ahler metric of matter fields among 3-point couplings.
In magnetized D-brane models, such a factor is universal, i.e.,
\begin{align}
K_{i\bar i}K_{j \bar j}K_{k\bar k}=(2 {\rm Im}\tau)^{-3/2},
\end{align}
if there are no corrections on the modular weights.

Suppose that the modular weights shift as
\begin{align}
k_i=-\frac12 + \Delta k_i, \quad k_j=-\frac12 + \Delta k_j, \quad k_k=-\frac12 + \Delta k_k,
\label{eq:shift-1}
\end{align}
for three matter fields in the above 3-point couplings.
However if the following relation: 
\begin{align}
\Delta k_ i + \Delta k_j+\Delta k_k = 0,
\label{eq:shift-2}
\end{align}
is satisfied, they lead to the equivalent low-energy effective field theory with the same physical couplings 
\footnote{See e.g. Refs. \cite{Cremades:2004wa,Kobayashi:2016ovu}.}.

This discussion can be easily extended to generic $n$-points couplings, 
where matter fields have the modular weights $k_i = -1/2 +\Delta k_i$ for $(i=1,\cdots, n)$.
If such corrections satisfy 
\begin{align}
\sum_{i=1}^n \Delta k_i = 0,
\end{align}
physical couplings are equivalent between the effective theories with and without corrections.
That suggests that if corrections $\Delta k_i$ are related to a symmetry and  
$\Delta k_i$ are proportional to its charges, its invariance would lead to 
the above situation.

\section{Modular weight corrections}

\subsection{Intersecting D-brane models}
\label{sec:typeIIA}

In type IIA intersecting D-brane models, 
the modular weights were also studied 
in Refs.~\cite{Blumenhagen:2006ci,Kors:2003wf,Lust:2004cx,Akerblom:2007uc,Blumenhagen:2007ip}, 
and they depend on intersecting angles of D-branes.
Similar to the previous sections, we concentrate one $T^2$ among $T^2 \times T^2 \times T^2$, 
and we denote its K\"ahler modulus $T$, which corresponds to $U={\rm Im}\tau + i{\rm Re}\tau$ 
in magnetized D-brane models.
For example, we use the Ansatz in Refs.~\cite{Akerblom:2007uc,Blumenhagen:2007ip},
where the K\"ahler potential and the modular weight are written by
\begin{align}
K=(T + \bar T)^{k_{ab}} |\phi_{ab}|^2, \qquad 
k_{ab}=-\frac 12 \pm {\rm sign}(I_{ab})\frac{\theta_{ab}}{2}.
\label{eq:int-weight}
\end{align}
Here, $I_{ab}$ is the intersecting number between D$_a$-brane and D$_b$-brane, and 
$\theta_{ab}$ denotes the angle $\pi \theta_{ab}$ of their intersecting point,
where the open string corresponding to the matter field $\phi_{ab}$ is localized.
We choose the angle region as $-1< \theta_{ab}<1$, and the sign such that $-1<k<0$ \cite{Akerblom:2007uc,Blumenhagen:2007ip}.
Similar results were obtained for twisted sectors in heterotic orbifold models 
\cite{Dixon:1989fj,Kaplunovsky:1995jw,Ibanez:1992hc,Ibanez:2012zz}.

The modular weights of matter fields are not universal because of the shifts proportional to the intersecting angles.
However, three matter fields localized at the intersecting points D$_a$-D$_b$, D$_b$-D$_c$, and D$_c$-D$_a$ 
are allowed to couple to each other if the following condition:  
\begin{align}
{\rm sign}(I_{ab})\theta_{ab} + {\rm sign}(I_{bc})\theta_{bc} + {\rm sign}(I_{ca})\theta_{ca} =0,
\end{align}
is satisfied.
This relation corresponds to Eqs.~(\ref{eq:shift-1}) and (\ref{eq:shift-2}).
Thus, physical couplings in the 4D effective field theory are equivalent between the modular weights 
with and without the corrections depending on intersecting angles.
Here, the rotation symmetry and the (twisted) boundary conditions seem to be important 
and it requires the sum of intersecting angles to vanish in allowed couplings.

We could discuss the T-dual of the above discussion.
Then, we may realize the same situation in type IIB magnetized D-brane models, where 
the $U(1)$ gauge symmetry associated with $U(1)$ magnetic fluxes determines the selection rule of allowed couplings. 
In the next subsection, we study another correction.

\subsection{Singular gauge transformation}
\label{sec:singular}

Here, we discuss another plausible origin of the modular weight shift $\Delta k_i$ in 
magnetized D-brane models\footnote{
One possibility would be corrections from the Dirac-Born-Infeld action.
Such corrections on the K\"ahler metric were computed in Ref.~\cite{Abe:2021uxb}, but it is not clear about 
the deviation of modular weights.}.

We start with a brief review of blowing-up orbifolds \cite{Kobayashi:2019fma,Kobayashi:2019gyl}.
The orbifold has singular points.
We can blow-up it to a smooth manifold.
For example, we replace singular points and their nearby regions of the $T^2/Z_N$ orbifold fixed points by parts of $S^2$.\footnote{See for the blow-up of higher-dimensional orbifolds, e.g., Refs. \cite{GrootNibbelink:2007lua,Leung:2019oln}.}
Wave functions on such blow-up $T^2/Z_N$ orbifolds were studied in Refs.~\cite{Kobayashi:2019fma,Kobayashi:2019gyl}.
On the $T^2/Z_N$ orbifold, wave functions have twisted boundary conditions (\ref{eq:twist-BC}).
Such a phase must be removed on a smooth manifold.
Indeed, we can remove such a phase by a singular gauge transformation.
The singular gauge transformation was studied for wave functions on magnetized $S^2$ with vortexes in Ref.~\cite{Dolan:2020sjq}. 
Localized fluxes and singular gauge transformation were also 
studied on the $T^2/Z_2$ orbifold in Refs. \cite{Buchmuller:2015eya,Buchmuller:2018lkz}. 
Then it was applied to blow-up of orbifolds in Refs.~\cite{Kobayashi:2022tti,Kobayashi:2022xsk}, 
where the following gauge transformation was studied
\begin{align}
A \to A + \delta A, \quad \delta A =iU_{\xi^F} d U^{-1}_{\xi^F},
\end{align}
with
\begin{align}
U_{\xi^F}=g(z)^{\xi^{F}/2}\overline g(z)^{-\xi^F/2}, \quad g(z)=e^{\frac{\pi}{2 {\rm Im}\tau}z^2}
  \vartheta \begin{bmatrix}
    \frac12 \\
    -\frac12 \\
  \end{bmatrix}
  (z,\tau) .
\end{align}
We study the wave function around the fixed point $z=0$.
Then, we find 
\begin{align}
 \delta A \simeq -i \frac12\frac{\xi^F}{z}dz+ i \frac12\frac{\xi^F}{\bar z}d\bar z.
\end{align}
That induces the modification of the field strength,
\begin{align}
\frac{\delta F}{2 \pi} = i\xi^F\delta( z) \delta (\bar z)dz \wedge d \bar z,
\end{align}
which is the localized flux at $z=0$.

Similarly, we perform the singular gauge transformation of the spin connection $\omega$,
\begin{align}
\delta \omega =iU_{\xi^R} d U^{-1}_{\xi^R} , \quad U_{\xi^R}=g(z)^{\xi^{R}/2}\overline g(z)^{-\xi^R/2},
\end{align}
where $\xi^R$ corresponds to the localized curvature.
Since the spin connection on $S^2$ is the same form as the gauge potential \cite{Conlon:2008qi}, wave functions on the $T^2/Z_N$ orbifold can be written by 
\begin{align}
\tilde \psi^{j,M}_m(z,\tau) = U_{\xi^F}^qU^{-s}_{\xi^R} \psi^{j,M}_m(z,\tau) =
g(z)^{q\xi^F/2 - s\xi^{R}/2}\overline g(z)^{-q\xi^F/2+s\xi^{R}/2} \psi^{j,M}_m(z,\tau),
\end{align}
where $q$ is the $U(1)$ charge of matter field and $s$ denotes its spin.

In Refs.~\cite{Kobayashi:2022tti,Kobayashi:2022xsk}, the singular gauge transformation was studied in the context of blowing-up orbifolds. 
In what follows, we examine the modular weights by focusing on the singular gauge transformation due to the localized flux as well as the localized curvature. Note that the following analysis is independent of whether we blow-up orbifolds or not.

Let us study the behavior of the new wave functions under the $S$-transformation.
The function $g(z)$ transforms 
\begin{align}
g(z) \to (-\tau)^{1/2}e^{3\pi i/4}g(z),
\end{align}
under the $S$-transformation.
In addition, by setting the normalization condition of $\tilde \psi^{j,M}_m(z,\tau)$ as
\begin{align}
  &\int d^2z ~\tilde \psi^{i,M}_m(z,\tau) \left(\tilde \psi^{j,M}_m(z,\tau)\right)^* = (2{\rm Im}\tau)^{-1/2-\Delta k}\delta_{i,j}, \label{eq:NormalizationPSItilde}
\end{align}
with $\Delta k = q\xi^F/2 \mp s\xi^{R}/2$,
the new wave functions transform 
\begin{align}
\tilde \psi^{i,M}(z,\tau) \to (-\tau)^{1/2+\Delta k} \sum_j \tilde \rho(S)_{ij} \tilde \psi^{j,M}(z,\tau),
\end{align}
under the $S$-transformation, where $\tilde \rho(S)_{ij}= e^{3\pi i \Delta k/2} \rho(S)_{ij}$.
Then, the modular weight of the above wave function $\tilde \psi^{j,M}_m(z,\tau) $ shifts 
from one of $\psi^{j,M}_m(z,\tau)$ by $\Delta k = q\xi^F/2 - s\xi^{R}/2$.

First, we study the shift of modular weights due to only the localized flux.
We consider three fields with charges $q_i$, $q_j$, and $q_k$.
Their modular weights shift as 
\begin{align}
\Delta k_i = q_i\frac{\xi^F}{2}, \quad \Delta k_j = q_j\frac{\xi^F}{2}, \quad \Delta k_k = q_k\frac{\xi^F}{2}.
\end{align}
If their Yukawa coupling is allowed, 
the gauge invariance requires $q_i+q_j+q_k=0$, which is nothing but Eqs.~(\ref{eq:shift-1}) and (\ref{eq:shift-2}).
Note that we can introduce the localized flux at the orbifold fixed points independently of whether we blow-up orbifolds or not.
Indeed, it can be induced by radiative corrections even if 
it vanishes at the tree level \cite{Lee:2003mc,Abe:2020vmv}, 
and it corresponds to the Fayet-Iliopoulos terms.

Remarkably, the form of this shift looks similar to one in intersecting D-brane models by the 
following replacement:
\begin{align}
\Delta k_i = q_i\frac{\xi^F}{2} \to \pm {\rm sign}(I_{ab})\frac{\theta_{ab}}{2}.
\end{align}
It is not clear whether this is accidental or has an underlying reason.
One plausible reason is as follows.
The radiative corrections induce the Fayet-Iliopoulos terms at the orbifold fixed points, 
which correspond to localized fluxes.
That is also relevant to U(1) anomalies.
Such anomalies can be canceled by the Green-Schwarz mechanism.
The Green-Schwarz anomaly cancellation mechanism requires a certain form of 
gauge kinetic functions. 
Indeed, the one-loop threshold corrections to the gauge kinetic function of the gauge group $G_A$ involve
\begin{align}
    (16\pi^2)f_{\rm 1-loop} =  -\left( C(G_A) - \sum_i T(R_i)\right) K_0 + 2\sum_{i} T(R_i) \log \det K_{ij},
\end{align}
with $K_0= -\ln (2 {\rm Im}\tau)$ and $K_{i\Bar{j}}=\partial_i \partial_{\Bar{j}} K$ being the moduli K\"ahler potential and matter K\"ahler metric (\ref{eq:int-weight}), respectively. Here, $C(G_A)$ and $T(R_i)$ denote the quadratic Casimirs in the adjoint of gauge symmetry $G_A$ and in the representation $R_i$ under which matter fields transform. By adopting the Ansatz (\ref{eq:int-weight}), the anomaly coefficients are extracted as \cite{Derendinger:1991hq}
\begin{align}
    b_A = -C(G_A) + \sum_i T(R_i) \left( 1+ 2k_{ii}\right),
\end{align}
with $k_{ii} = -1/2 + q_i \xi_F/2$. 
Since the tree-level gauge kinetic function of the gauge group $G_A$ depends on other moduli $T$ (K\"ahler moduli), the Green-Schwarz anomaly would be canceled by the transformation of other moduli \cite{Derendinger:1991hq,Ibanez:1992hc}: 
\begin{align}
    T \rightarrow T + \frac{\xi_F}{8\pi^2} \ln (c\tau +b),
\end{align}
up to an ${\cal O}(1)$ numerical coefficient.
We would study further elsewhere to understand this issue. 
The relation between localized fluxes and Green-Schwarz anomaly cancellation mechanism
were studied on the $T^2/Z_2$ orbifold in Ref. \cite{Buchmuller:2015eya}.

Next, we study the shift of modular weights due to the localized curvature.
When we compute the 3-point couplings among modes with the spins, 0 and $1/2$, $-1/2$, 
which corresponds to the Yukawa couplings among two spinors and scalar, 
the corrections due to the localized curvature cancel each other 
in the sum like $\Delta k_i + \Delta k_j +\Delta k_k$.
We have the same results for the 3-point couplings among two spinors and the vector along the compact direction. 
These behaviors look similar to the above. 
However, for the localized curvature, bosons and fermions have different modular weight shifts, which will be discussed in more detail in Appendix \ref{app:modular_weight}.
In a SUSY theory, we may consider the singular gauge transformation due to only the 
localized flux.
If we include the correction due to the localized curvature, 
we may need to introduce another U(1) transformation to cancel 
the difference by spins e.g., $U(1)_R$ symmetry. 
This mechanism would correspond to the Scherk-Schwarz compactification where 
the boundary conditions of scalar and fermions are different to each other by the $U(1)_R$ twistings. 
A detailed phenomenology of such a SUSY-breaking compactification will be left for future work.

We have studied the wave functions on the orbifolds.
Similarly, if it is possible to introduce the localized flux on $T^2$, 
we can study the wave functions $\psi^{j,M}(z,\tau)$ on $T^2$.
Their boundary conditions around $z=0$ have no phase.
However, we perform a singular gauge transformation by $U_{\xi ^F}^q$.
Then, they have twisted boundary conditions due to the localized flux, 
and their modular weight shift from $1/2$ by $\Delta k = q\xi^F/2$ in a way similar to the previous discussions, although the Fayet-Iliopoulos term can not appear on $T^2$.

\section{Soft SUSY breaking terms}
\label{sec:SUSY}

Next, we study soft SUSY breaking terms.
In section \ref{sec:LEEFT}, we mentioned that the normalization of kinetic term is not physically observable.
However, if we treat the moduli as not constants, but dynamical fields, 
the K\"ahler metric provides us with couplings between matter fields and moduli fields.
Such couplings are important when SUSY is broken by moduli $F$-terms with other sources.
Now, we study the so-called moduli-mediation of SUSY breaking 
\cite{Kaplunovsky:1993rd,Brignole:1993dj,Kobayashi:1994eh,Brignole:1995fb,Ibanez:1998rf}.
Sfermion masses can be written by \cite{Kaplunovsky:1993rd}
\begin{eqnarray}
 m_i^2= m_{3/2}^2-\sum_{X} |F^X|^2 \partial_X \partial_{\bar X}\ln K_{i \bar i}\,,
\end{eqnarray}
in the canonically normalized basis, where $X$ denotes the sources of SUSY breaking.
Similarly, the so-called $A$-term in SUSY breaking Lagrangian,
\begin{align}
{\cal L}_{\rm A} = h_{ijk}\phi^i\phi^j\phi^k + \cdots, 
\end{align}
can be written by \cite{Kaplunovsky:1993rd}
\begin{eqnarray}
A_{ijk} =A_i+A_j+A_k -\sum_{X} \frac{F^X}{Y_{ijk}} \partial_X Y_{ijk}\,, 
\end{eqnarray}
with 
\begin{eqnarray}
A_i = \sum_X F^X \partial_X \ln e^{-K_0/3}K_{i\bar i}\,,
\end{eqnarray}
where we used the convention $h_{ijk}=Y_{ijk}A_{ijk}$, and 
$K_0$ denotes the K\"ahler potential of the modulus, e.g. 
\begin{align}
K_0= - \ln (U + \bar U),
\end{align}
for type IIB magnetized D-brane models, and 
\begin{align}
K_0= - \ln (T + \bar T),
\end{align}
for type IIA intersecting D-brane models.

For simplicity, we assume that the $F$-term of the modulus $U$ corresponding to one of $T^2$ among 
$T^2 \times T^2 \times T^2$ develops its vacuum expectation value and contributes SUSY breaking. \footnote{See, Ref. \cite{Kikuchi:2022pkd}, for the modular symmetry of soft SUSY-breaking terms.}
Then, sfermion masses can be written by 
\begin{eqnarray}
 m_i^2= m_{3/2}^2+k_i \frac{|F^U|^2}{(U+\bar U)^2} \,,
\end{eqnarray}
for the matter fields with the modular weight $k_i$.
Thus, the difference between the modular weights affects sfermion masses, which can be 
observables.
If we could observe deviations of sfermion masses from the universal value $m_{3/2}$ by experiments, 
we would know the ratio among modular weights.
Hence, the corrections of modular weights are important in sfermion masses.

On the other hand, the $A$-term coefficients can be written by 
\begin{eqnarray}
A_{ijk}&=&A_{ijk}^0+A'_{ijk}, \nonumber \\
A_{ijk}^0&=& (k_i+k_j+k_k -1)\frac{ F^U}{(U + \bar U)}, \qquad 
A'_{ijk}=-\frac{F^U}{Y_{ijk}}\frac{dY_{ijk}}{d U} \, .
\label{Aterm}
\end{eqnarray}
Only the combination of modular weights, $(k_i+k_j+k_k )$, appears, and 
it is obtained as $(k_i+k_j+k_k )=-3/2$ in both magnetized D-brane models and intersecting D-brane models.
Even if we shift the modular weights like Eq.(\ref{eq:shift-1}) and Eq.(\ref{eq:shift-2}), 
the $A$-term has no effect.
Similarly, the sum of sfermion masses squared is universal,
\begin{align}
m^2_i + m^2_j + m^2_k = 3m^2_{3/2} +(k_i+k_j+k_k)\frac{|F^U|^2}{(U+\bar U)^2} 
= 3m^2_{3/2} -\frac32 \frac{|F^U|^2}{(U+\bar U)^2} .
\end{align}
Renormalization group equations of sfermion masses include only $A_{ijk}$ and the sum $(m^2_i + m^2_j + m^2_k)$.
Thus radiative corrections have no effect even if  
we shift the modular weights like Eq.(\ref{eq:shift-1}) and Eq.(\ref{eq:shift-2}).
(See for implications on renormalization group equations, e.g., Refs.~\cite{Kawamura:1997cw,Kobayashi:1997qx}.)

We have discussed the $F$-term of a single modulus for simplicity.
However, a generic situation is more complicated on $T^2 \times T^2 \times T^2$.
In magnetized D-brane models, we denote the modulus $U_\alpha$ and its $F$-term $F^{U_\alpha}$ corresponding to the $\alpha$-th $T^2$ 
among $T^2 \times T^2 \times T^2$.
Then, sfermion masses can be written by 
\begin{eqnarray}
 m_i^2= m_{3/2}^2+\sum_\alpha k_i^\alpha \frac{|F^{U_\alpha}|^2}{(U_\alpha +\bar U_\alpha)^2} = 
m_{3/2}^2-\frac12 \sum_\alpha  \frac{|F^{U_\alpha}|^2}{(U_\alpha +\bar U_\alpha)^2} +\sum_\alpha \Delta k_i^\alpha \frac{|F^{U_\alpha}|^2}{(U_\alpha +\bar U_\alpha)^2}\,,
\end{eqnarray}
in magnetized D-brane models.

Furthermore, we give a comment on $D$-term contributions.
In general, sfermion masses have $D$-term contributions.
(See, Refs.~\cite{Kawamura:1996wn,Kawamura:1996bd,Higaki:2003ig}, for $D$-term contributions including anomalous U(1) symmetry in the moduli-mediation.)
For simplicity, we assume the existence of extra U(1) symmetry and it is broken by some mechanism. 
That leads to the non-vanishing $D$-term, which leads to additional effects on sfermion masses, 
\begin{align}
\Delta m_i^2 = q_i D,
\end{align}
where $q_i$ denotes U(1) charge of the matter field.
Similarly, we have additional effects due to $D$-terms of non-Abelian gauge symmetries.
Thus, prediction on sfermion masses becomes complicated.
However, U(1) charges must be conserved in allowed Yukawa couplings.
Hence, the $D$-term has no effect on the sum $(m^2_i + m_j^2 +m_k^2)$.
Also, it is impressive that 
the deviations from the universal mass $m_0$, 
\begin{align}
m_0^2 = m_{3/2}^2-\frac12 \sum_\alpha  \frac{|F^{U_\alpha}|^2}{(U_\alpha +\bar U_\alpha)^2} 
\end{align}
are proportional to the charge $q_i$ and $\Delta k_i^\alpha$, which is also proportional to the charge 
corresponding to the localized flux and the $Z_N$ charge of the twisted boundary condition.

 \section{Discussion on massive modes}
\label{sec:massive}

Here, we discuss more about the modular weights including oscillating modes and massive modes.
In heterotic orbifold models, oscillated twisted modes are created by $\alpha_{-1}$ from the twisted ground state.
When the twisted ground state has the modular weight $k$, 
the modular weight of a singly oscillated twisted mode is equal to $k-1$.
We study whether there is a similar behavior in magnetized D-brane models.

The mass level of magnetized D-brane models is similar to the harmonic oscillator 
\cite{Cremades:2004wa,Berasaluce-Gonzalez:2012abm,Hamada:2012wj}, 
as shown in Appendix \ref{app:wave}.
We can create massive modes by operating $\nabla_z$ on the massless mode $\psi^{j,M}$.
The operator $ \nabla_z$ transforms 
\begin{align}
 \nabla_z \to (-\tau) \nabla_z,
\end{align}
under the $S$-transformation.
That is, the operator  $\nabla_z$ has the modular weight 1.
Note that the modular weight of  4D matter fields has the opposite sign of the modular weight of 
wave function.
Then, the 4D first massive fields have $k-1$ when the massless mode has the modular weight $k$.
This behavior is the same as one in heterotic twisted modes.
Thus, in magnetized D-brane models, the wave functions including the $n$-th massive modes created by  $(\nabla_z)^n$ have the modular weights,
\begin{align}
k^{(n)}=\frac{\cal A}{4 \pi M}\lambda_n=n + \frac12,
\end{align}
when there is no correction $\Delta k$.
This is nothing but the energy spectrum of the harmonic oscillator and the eigenvalues $\lambda_n$ of  Laplacian $\Delta$ up to a constant.
(See for details of massive modes Appendix \ref{app:wave}.)
That also suggests that the zero-point energy is equal to $1/2$, which corresponds to the modular weight 
of the zero-mode wave functions and corresponds to the modular weight $-1/2$ of 
4D matter field.

In heterotic string theory, the K\"ahler metric and the modular weight are computed by 
string amplitude within the framework of world-sheet conformal field theory.
The operator $i\partial X$ corresponding to $\alpha_{-1}$ increases the world-sheet conformal dimension $h$ by 1.
The quantum system of magnetized models on $T^2$ has the $OSp(1|2)$ subalgebra of superconformal 
algebra (super-Virasoro algebra) including $\nabla_z$, which increases $k^{(n)}$ by 1.
We define 
\begin{align}
\ell_0 = \frac{\cal A}{8 \pi M}\Delta, 
\qquad g_{-1/2}=\frac{1}{\sqrt 2} \nabla_z, \qquad 
g_{1/2}=\frac{1}{\sqrt 2} \nabla_{\bar z}, 
\qquad \ell_{\pm 1} = (g_{\pm 1/2})^2 .
\end{align}
They satisfy the following algebraic relations:
\begin{align}
[\ell_m,\ell_n]=(m-n)\ell_{m+n}, \qquad 
[\ell_m, g_r]=\left(\frac{m}{2} -r\right) g_r,
\qquad \{ g_r, g_s\} = 2\ell_{r+s},
\end{align}
where $m,n=0, \pm 1$ and $r,s= \pm 1/2$.
This is the $OSp(1|2)$ subalgebra of super-Virasoro algebra.
This correspondence leads the correspondence between the conformal dimension $h$ 
of the world-sheet conformal field theory in heterotic string theory as well as 
intersecting D-brane models and the eigenvalues of Laplacian in 
magnetized $T^2$ models.
That is, the eigenvalues $d$ of $\ell_0$ are related with the levels of Laplacian, which correspond to
the energy spectrum of harmonic oscillator, as 
\begin{align}
2d =  k^{(n)} .
\end{align}
In this correspondence of algebras, 
 the operator $i \partial X$ corresponding to $\alpha_{-1}$ has the world-sheet conformal 
dimension $h=1$, and the modular weight in heterotic orbifold models is 
decreased by $\alpha_{-1}$.
On the other hand, the operator $\nabla_z$ has the eigenvalue $d=1/2$ of $\ell_0$, which increases 
$k^{(n)}$ by 1, and decreases the modular weight of the 4D matter fields by $-1$.

In this discussion, the zero-point energy $1/2$ in the harmonic oscillator corresponds to the modular weight of the zero-mode wave functions.
The modular weight $1/2$ would be natural for spinor fields, because theses are the spinor representation 
of $SO(2)$ on $T^2$ and $S^2$ is the $\pi$ rotation, i.e. $S^2= -1$.

Now, let us consider the harmonic oscillators with ``twisted boundary conditions''.
We define them as the creation and the annihilation operators of the twisted string.
They satisfy 
\begin{align}
[\tilde \alpha_{m+\theta}, \alpha_{n-\theta} ] = \delta_{m, -n}(m+\theta).
\end{align}
For such an oscillator, the zero point energy shifts by $\theta/2$.

\section{Conclusion}
\label{sec:Conclusion}

We have studied the modular weights of type IIB magnetized D-brane models.
The modular weights have corrections by the localized flux as well as the localized curvature.
On the other hand, corrections of modular weights were discussed in type IIA intersecting D-brane models 
through the analysis on one-loop gauge corrections.
Similar corrections appear in type IIB magnetized D-brane models, too.
As a result, the form of corrected modular weights looks similar.

Such corrections do not affect physical couplings when we think that the K\"ahler metric is just a 
normalization factor of the kinetic term.
Cancellation of modular weight corrections in physical couplings is originated by the underlying symmetry 
such as $Z_N$ rotation symmetry and gauge invariance.
Corrected modular weights lead to different behaviors when we think that 
the K\"ahler metric is a coupling between the modulus and matter fields.
One example is SUSY-breaking sfermion masses, which depend on the modular weights, 
although the $A$-term coefficients $A_{ijk}$ and the sum of sfermion masses squared seem to be 
the same between two models with and without corrections.

We have also studied the modular weights of massive wave functions in magnetized D-brane models.
Its increasing behavior is similar to oscillated twisted modes in heterotic orbifold models.
This result is originated from the $OSp(1|2)$ subalgebra of super-Virasoro algebara.
This coincidence suggests that the zero point energy $1/2$ corresponds to the 
modular weights of massless wave functions 1/2 and the 4D massless matter fields $-1/2$ 
in magnetized D-brane models without the above corrections.

Although we found that the coincidence of the modular weight shifts between 
magnetized D-brane models and intersecting D-brane models, 
its underlying reason is unclear.
In the intersecting D-brane side, it follows the Ansatz on consistency of one-loop gauge coupling corrections, 
while the shift can be originated from the localized flux as well as the localized curvature in its T-dual magnetized D-brane side. 
If the localized flux is induced by radiative corrections, this coincidence may have an underlying reason related 
gauge anomalies and the Green-Schwarz mechanism.
Alternatively, the modular weight shifts may be related to some symmetry such as gauge invariance and 
the rotation symmetry in order to lead to the same physical couplings.
That is, the shifts are determined by the charges of those symmetries to control the coupling selection rules.
This aspect may be the key to understand the coincidence.
At this stage, the true reason is unclear.
It is important to study further in order to understand the reason of our results.
We would study it elsewhere.

\vspace{1.5 cm}
\noindent
{\large\bf Acknowledgments}\\

This work was supported by JPSP KAKENHI Grant Numbers JP20K14477(HO) and 
JP20J20388(HU) and JP22J10172(SK), and 
JST SPRING Grant Number JPMJSP2119(KN).



\appendix

\section{Matter wave functions}
\label{app:wave}

\subsection{Zero modes}

Here, we briefly review the zero-mode wave functions on the magnetized $T^2$\cite{Cremades:2004wa}. We do not take into account the Wilson lines or the Scherk-Schwarz phases. The two dimensional torus $T^2$ is defined as the quotient of a complex plane $\mathbb{C}$ by a lattice $\Lambda$ spanned by two vectors $e_1, e_2 \in \mathbb{C}$, i.e. $T^2 \simeq \mathbb{C}/\Lambda$. We define the complex coordinate of $T^2$ by
$z=u/e_1$ where $u$ is that of the complex plane $\mathbb{C}$. Thus, we have identifications, $z \sim z+1$ and $z \sim z + \tau$.
The complex structure modulus of $T^2$ is given by $\tau=e_2/e_1, ({\rm Im}\tau > 0)$. The metric of $T^2$ is written as 
\begin{equation}
    ds^2 = 2 h_{\mu \bar{\nu}} dz^{\mu} d\bar{z}^{\bar{\nu}},
\end{equation}
where
\begin{equation}
    h = |e_1|^2
    \begin{pmatrix}
    0 & 1/2 \\
    1/2 & 0
    \end{pmatrix}.
\end{equation}
The magnetic flux background in Eq. (\ref{eq:magne}) corresponds to the vector potential 
\begin{equation}
    A(z) = \frac{\pi M}{{\rm Im}\tau} {\rm Im}(\bar{z} dz),
\end{equation}
which satisfies the following boundary conditions:
\begin{align}
\begin{aligned}  
A(z+1) &= A(z) + d\left( \frac{\pi M}{{\rm Im}\tau} {\rm Im}z    \right) = A(z) + d\chi_1(z), \\
 A(z+\tau) &= A(z) + d\left( \frac{\pi M}{{\rm Im}\tau} {\rm Im}(\bar{\tau}z)    \right) = A(z) + d\chi_2(z).
    \end{aligned}
\end{align}
We consider the two dimensional spinor on $T^2$,
\begin{equation}
    \Psi(z,\tau) = \begin{pmatrix}
    \psi_+ \\
    \psi_-
    \end{pmatrix},
\end{equation}
which couples to $A(z)$ with $U(1)$ unit charge $q=1$. Here, $\pm$ denotes the chirality. The spinor $\Psi$ satisfies
\begin{align}
\begin{aligned}
\label{eq: boundary_conditions}
\Psi(z + 1,\tau) &= e^{i \chi_1(z)}\Psi(z,\tau)
    = e^{\frac{\pi i M}{{\rm Im}\tau} {\rm Im}z} \Psi(z,\tau), \\
\Psi(z + \tau,\tau) &= e^{i \chi_2(z)}\Psi(z,\tau) 
= e^{\frac{\pi i M}{{\rm Im}\tau} {\rm Im}(\bar{\tau}z)  } \Psi(z,\tau) ,
\end{aligned}
\end{align}
under the translations. For the consistency of boundary conditions along the contractible loop in $T^2$, the magnetic flux $M$ must be an integer. This is referred to as Dirac's quantization condition.

Now, we consider the Dirac equation for $\Psi$. We take the following gamma matrices 
\begin{equation}
\Gamma^z = \frac{1}{e_1}
    \begin{pmatrix}
    0 & 2 \\ 0 & 0
    \end{pmatrix},\ 
\Gamma^{\bar{z}} = \frac{1}{\bar{e}_1}
    \begin{pmatrix}
    0 & 0 \\ 2 & 0
    \end{pmatrix},   \label{eq:gammamatrix}
\end{equation}
which satisfy $\{ \Gamma^z , \Gamma^{\bar{z}} \} = 2 h^{z\bar{z}}$. 
The covariant derivatives are given by 
\begin{equation}
    \nabla_z = \partial - i A_z, \quad
    \nabla_{\bar{z}} = \bar{\partial} - i A_{\bar{z}}.
\end{equation}
Then, we can write down the Dirac operator as
\begin{equation}
    i \slash{D} = i \Gamma^z \nabla_z + i \Gamma^{\bar{z}} \nabla_{\bar{z}} 
    = i 
    \begin{pmatrix}
    0 & \frac{2}{e_1} \left( \partial - \frac{\pi M}{2 {\rm Im}\tau}\bar{z} \right)
\\
\frac{2}{\bar{e}_1} \left(  \bar{\partial} + \frac{\pi M}{2 {\rm Im}\tau}{z}
\right) & 0
    \end{pmatrix} 
\equiv  i
\begin{pmatrix}
0 & - D^{\dagger} \\
D & 0
\end{pmatrix}.
\end{equation}
We are interested in the zero-mode wave functions satisfying $i \slash{D}\Psi = 0$ which is equivalent to 
\begin{align}
\label{eq: zero-mode_eom_+}
    D \psi_+(z,\tau) &= 0, \\
\label{eq: zero-mode_eom_-}
    D^{\dagger} \psi_-(z,\tau) &=0.
\end{align}
Solving Eq. (\ref{eq: zero-mode_eom_+}) under the boundary conditions Eq.(\ref{eq: boundary_conditions}) shows that the zero-mode $\psi_+$ has degeneracy number $M$ and they are written as
\begin{equation}
\label{eq: psi_+}
    \psi_+^{j,M}(z,\tau) = \frac{M^{1/4}}{\mathcal{A}^{1/2}}  e^{\pi i M z \frac{{\rm Im}z}{{\rm Im}\tau}}\ \vartheta
     \begin{bmatrix}
         \frac{j}{M} \\ 0
     \end{bmatrix}(Mz, M\tau).
\end{equation}
Here, $\vartheta$ is the Jacobi theta function as defined in Eq.(\ref{eq:theta-f}). Notice that $\psi_+$ is normalizable only if $M>0$.

On the other hand, the negative chirality zero-mode $\psi_-(z,\tau)$ is normalizable only if $M<0$ and its degeneracy is  $|M|$,  
\begin{equation}
\label{eq: psi_-}
      \psi_-^{j,M}(\bar{z},\bar{\tau}) = \frac{|M|^{1/4}}{\mathcal{A}^{1/2}}  e^{\pi i M \bar{z} \frac{{\rm Im}\bar{z}}{{\rm Im}\bar{\tau}}}\ \vartheta
     \begin{bmatrix}
         \frac{j}{M} \\ 0
     \end{bmatrix}(M\bar{z}, M\bar{\tau}),
\end{equation}
where $j = 0,1,...,|M|-1$.

\subsection{Massive modes}
Here, we review the fermion massive modes $\Psi_n$ \cite{Cremades:2004wa,Berasaluce-Gonzalez:2012abm,Hamada:2012wj}. The corresponding Dirac equation is \begin{equation}
    i\slash{D}\Psi_n(z,\tau) = m_n \Psi_n({z,\tau}),
\end{equation}
with $m_n > 0, (n=1,2,\cdots)$. By acting $i\slash{D}$ one more time, we find that wave functions of the massive modes satisfy
\begin{equation}
 (i \slash{D})^2 \Psi_{n} = 
    \begin{pmatrix}
    D^{\dagger} D & 0 \\
    0 & D D^{\dagger}
    \end{pmatrix} 
    \begin{pmatrix}
    \psi_{+,n}\\
    \psi_{-,n}
    \end{pmatrix}
    =
    m_n^2
    \begin{pmatrix}
    \psi_{+,n}\\
    \psi_{-,n}
    \end{pmatrix}.    
\end{equation}
The Laplacian on $T^2$ is defined by
\begin{equation}
        \Delta = \frac{1}{2} \{D, D^{\dagger} \}.
\end{equation}
Then, we have the following relations
\begin{align}
    \begin{aligned}
 &\Delta = D^{\dagger}D + \frac{2 \pi M}{\mathcal{A}},\quad [D, D^{\dagger}] = \frac{4 \pi M}{\mathcal{A}}, \\
 &[\Delta, D^{\dagger}] = \frac{4 \pi M}{\mathcal{A}}D^{\dagger}, \quad
  [\Delta, D] = -\frac{4 \pi M}{\mathcal{A}}D.
    \end{aligned}
\end{align}
This is the algebra of the harmonic oscillator in quantum mechanics. If we normalize the creation and the annihilation operators by,
\begin{equation}
    a^{\dagger} = \sqrt{\frac{\mathcal{A}}{4 \pi |M|}} D^{\dagger}, \quad a = \sqrt{\frac{\mathcal{A}}{4 \pi |M|}}D,
\end{equation}
they satisfy $[a,a^{\dagger}] = \pm1$ for ${\rm sign}(M)=\pm1$ respectively. 
We notice that  the zero-mode wave functions $\psi_{\pm}^{j,M}$ in Eqs. (\ref{eq: psi_+}) and (\ref{eq: psi_-}) are the eigenstates of $\Delta$ with the smallest eigenvalue $2\pi |M|/\mathcal{A}$. Thus, by acting $a^{\dagger}$ ($a$) to $\psi_{+}^{j,M}$ ($\psi_{-}^{j,M}$), we obtain the massive excited modes,
\begin{equation}
\label{eq: massive_modes}
    \psi_{n, +}^{j,M} = \frac{1}{\sqrt{n!}} (a^{\dagger})^n \psi_{+}^{j,M}, \quad \psi_{n, -}^{j,M} = \frac{1}{\sqrt{n!}} (a)^n \psi_{-}^{j,M},
\end{equation}
whose eigenvalues of $\Delta$ are given by
\begin{equation}
    \lambda_n = \frac{2\pi |M|}{\mathcal{A}}(2n+1).
\end{equation}
Thus, the corresponding values of the mass squared are 
\begin{equation}
    m_n^2 = \frac{4 \pi |M|}{\mathcal{A}}n.
\end{equation}
Note that $\psi_{n,\pm}^{j,M}$ in Eq. (\ref{eq: massive_modes}) satisfy the boundary conditions Eq. (\ref{eq: boundary_conditions}) as required.

The derivative operators $\nabla_z$ and $\nabla_{\bar{z}}$ also satisfy the following relations, 
\begin{align}
\begin{aligned}
 &[\Delta, \nabla_z] = \frac{4 \pi M}{\mathcal{A}} \nabla_z, \quad
  [\Delta, \nabla_{\bar{z}}] = -\frac{4 \pi M}{\mathcal{A}} \nabla_{\bar{z}}.
\end{aligned}
\end{align}
Thus, massive modes can also be written by acting $\nabla_z$ and $\nabla_{\bar{z}}$ to the zero-mode wave functions. The positive chirality massive modes are 
\begin{equation}
    \psi_{n,+}^{j,M} \propto (\nabla_z)^n \psi_{+}^{j,M},
\end{equation}
whose eigenvalues of $\Delta$ are given by the above $\lambda_n$.
Similarly, the negative chirality massive modes are given by 
\begin{equation}
    \psi_{n,-}^{j,M} \propto (\nabla_{\bar{z}})^n \psi_{-}^{j,M}.
\end{equation}

\section{Modular weights of scalar and spinor fields}
\label{app:modular_weight}

The $Z_N$ twisted boundary conditions of scalar ($s=0$) and spinor ($s=\pm1/2$) fields on $T^2/Z_N$ orbifold are given by
\begin{align}
\psi_{s,m}^{j,M}(\theta z, \tau) = {\cal S}_s V_m \psi_{s,m}^{j,M}(z, \tau),
\end{align}
and they satisfy
\begin{align}
\psi_{s,m}^{j,M}(\theta^N z, \tau) = {\cal S}_s^N V_m^N \psi_{s,m}^{j,M}(z, \tau) = \psi_{s,m}^{j,M}(z, \tau).
\end{align}
First, let us consider the scalar mode.
In this case, ${\cal S}_0=1$ and then $V_m^N=1$ should be satisfied.
Thus, we can set
\begin{align}
V_m = e^{2\pi im/N} \quad (m \in \mathbb{Z}/N\mathbb{Z}).
\end{align}
that is, the $Z_N$ twisted boundary condition of the scalar field is written by
\begin{align}
\psi_{0,m}^{j,M}(\theta z, \tau) = e^{2\pi im/N} \psi_{0,m}^{j,M}(z, \tau). \label{eq:BCscalar}
\end{align}
Next, let us consider the spinor mode.
Here, we define
\begin{align}
{\cal S}_{|1/2|} \equiv
\begin{pmatrix}
{\cal S}_{+1/2} & 0 \\
0 & {\cal S}_{-1/2}
\end{pmatrix},
\end{align}
which acts on the spinor modes $(\psi_{+1/2,m}^{j,M}, \psi_{-1/2,m}^{j,M})^T$.
Under $Z_N$ twist, it should satisfy
\begin{align}
\theta \Gamma^z = {\cal S}_{|1/2|} \Gamma^z {\cal S}_{|1/2|}^{-1}, \quad
\theta^{-1} \Gamma^{\bar{z}} = {\cal S}_{|1/2|} \Gamma^{\bar{z}} {\cal S}_{|1/2|}^{-1},
\end{align}
where $\Gamma^z$ and $\Gamma^{\bar{z}}$ are defined in Eq.~(\ref{eq:gammamatrix}).
Then, ${\cal S}_{\pm1/2}$ can be expressed as
\begin{align}
{\cal S}_{\pm1/2} = \theta^{q'} \theta^{\mp1/2},
\end{align}
where there is a degree of freedom of $U(1)$ for a spinor field and $q'$ denotes the $U(1)$ charge.
Hereafter, we denote the $U(1)$ as $U(1)'$.
Moreover, ${\cal S}_{\pm1/2}$ should satisfy ${\cal S}_{\pm 1/2}^N = 1$, and then $q'$ must be
\begin{align} 
q' = 1/2 + m' \quad (m' \in \mathbb{Z}/N\mathbb{Z}),
\end{align}
(while $q'=0$ for a scalar field).
Thus, the $Z_N$ twisted boundary condition of the spinor field is written by
\begin{align}
\psi_{\pm1/2,m}^{j,M}(\theta z, \tau) = e^{2\pi i(m+q'\mp1/2)/N} \psi_{\pm1/2,m}^{j,M}(z, \tau). \label{eq:BCspinor}
\end{align}
Note that, when $M>0$, $\psi_{+1/2,m}^{j,M}(z, \tau)$ becomes the zero mode.
In addition, if it is also the superpartner of the scalar $\psi_{0,m}^{j,M}(z, \tau)$, $q'$ must be $q'=1/2$.


We consider the following singular gauge transformations,
\begin{align}
\tilde \psi^{j,M}_{s,m}(z,\tau) = U_{\xi^F}^q U_{\xi'^F}^{q'} U^{-s}_{\xi^R} \psi^{j,M}_{s,m}(z,\tau) ,
\end{align}
such that the $Z_N$ twisted boundary condition becomes
\begin{align}
\tilde \psi^{j,M}_{s,m}(\theta z,\tau) = \tilde \psi^{j,M}_{s,m}(z,\tau).
\end{align}
Here, $\xi^F$ and $\xi'^F$ denote the localized flux of $U(1)$ and $U(1)'$, while the localized curvature is $\xi^R=N-1$.
Since $U_{\xi}$ transforms under the $Z_N$ twist as $U_{\xi} \rightarrow \theta^{\xi} U_{\xi}$, the followings should be satisfied;
\begin{align}
&\tilde \psi^{j,M}_{s,m}(\theta z,\tau) = \theta^{m+q'-s+q\xi^F+q'\xi'^F-s\xi^R} \tilde \psi^{j,M}_{s,m}(z,\tau) = \tilde \psi^{j,M}_{s,m}(z,\tau), \\
\Leftrightarrow \ &
m+q\xi^F \equiv 0\ ({\rm mod}\ N), \quad \xi'^F \equiv \xi^R\ ({\rm mod}\ 2N).
\end{align}

By using the localized fluxes and the localized curvature, the modular weight is shifted from $1/2$ by $\Delta k = q\xi^F/2 + q' \xi'^F/2-s\xi^R/2$.
In particular, if they satisfy  $q'=s=+1/2$ and $\xi'^F=\xi^R$, 
the modular weights of both $\psi_{+1/2,m}^{j,M}(z, \tau)$ and $\psi_{0,m}^{j,M}(z, \tau)$ are 
the same, i.e., $1/2+\Delta k$ with $\Delta k = q\xi^F/2$.


\end{document}